\documentclass[11pt,final]{article}
\usepackage{graphicx,float,caption2}
\usepackage{hyperref}
\pdfoutput=1
\usepackage{setspace}          
\usepackage{amsmath,amssymb}   
\usepackage{wrapfig}
\usepackage{url}
\usepackage[super,comma,sort&compress]{natbib}
\usepackage{fancyhdr}
\usepackage{subfigure}

\title{Hydraulic Jump in a Non-wetting Film \\ Deposited inside of a Capillary}

\author{Alexandru Herescu\thanks{aherescu@mtu.edu} \hspace{0.6in} 
Jeffrey S. Allen\thanks{Corresponding author: jstallen@mtu.edu} \\
               Mechanical Engineering -- Engineering Mechanics \\
                Michigan Technological University \\
                }
     
\begin{document}

\date{}
\maketitle
\begin{abstract}
Fluid dynamics video
\end{abstract}

\section*{\emph{Experimental Observations}}
Film deposition experiments are performed in circular glass capillaries of 500 $\mu m$ diameter: the capillary is filled with water which is afterwards evacuated by nitrogen flowing into the tube. Two surface wettabilities are considered, contact angle $\theta$ = 30$^\circ$ (wetting) for water on glass and $\theta$ = 105$^\circ$ (non-wetting) when a hydrophobic coating is applied. As the meniscus translates a film is deposited on the inner wall of the capillary, which is seen as the bright region delimiting and surrounding the dark gas bubbles. Note that the film thickness varies with surface wettability, being thicker for $\theta$ = 105$^\circ$ as is apparent from the images showing the wetting and nonwetting deposition. It was observed that the liquid film deposited as the meniscus translates with a velocity $U$ presents a \emph{ridge} which also moves in the direction of the flow. \emph{The ridge} is bound by a contact line (at the receding end) moving at a velocity $U_{CL}$ and it translates over the deposited film. Movies recorded at 500 fps show a liquid column (bright) followed by a bubble train (dark) that forms as the gas displaces the liquid. At the end of the movie one can observe the film deposition process as well as the bubble formation due to the growing ridge, which eventually forms a plug. In the beginning it can be seen a film thicker than the Bretherton~\cite{Bretherton1961} prediction being deposited. If a certain Capillary number is exceeded, $\textbf{Ca}^*$ = 0.0034 in present experiments, the thick film ($\approx$ 20 $\mu m$) transitions through a \emph{hydraulic jump} into a thin film ($\approx$ 10 $\mu m$) which merges with the meniscus. The movie shows the deposition process at $\textbf{Ca}$ = 0.0038 $\ge \textbf{Ca}^*$, with a visible \emph{hydraulic jump}. For $\textbf{Ca} \le \textbf{Ca}^*$ (not shown here) the nonwetting film is also significantly thicker than the wetting film but does not present a jump. \\

\section*{\emph{Discussion}}
 The behavior of the ridge presents striking dissimilarities when the wettability is changed, $U_{CL}$ is significantly larger for the non-wetting case at the same capillary number $\textbf{Ca}$. Classical film deposition theory does not account for the existence of a contact line and it assumes perfect wetting. In contrast, \emph{the contact line dynamics fundamentally alter the deposition physics by causing the film to be non-stagnant}. As a consequence the non-wetting film is significantly thicker than Bretherton's prediction as is apparent from the images acquired at the same $\textbf{Ca}$ in both wetting and nonwetting test sections. Taylor bubbles also form due to the growth of the ridge and are differentiated by wettability, being much shorter and presenting a thicker film in the nonwetting case. The dynamics of the deposition is studied experimentally and it was observed that below $\textbf{Ca}^*$ = 0.0034 a thick film ($\approx$ 20 $\mu m$) is deposited, while if $\textbf{Ca}^*$ is exceeded a thick ($\approx$ 20 - 30 $\mu m$) and a thin film ($\approx$ 10 $\mu m$) coexist. The thick film adjacent to the ridge and the thin film are separated through a hydraulic jump. The hydraulic jump cannot be explained by the Froude condition of shock formation in shallow waters due to the insignificance of gravity in the case of a 10 $\mu m$ thick film. The shock formation is due to a \emph{capillary drainage mechanism} made possible by the presence of the moving contact line, and by the dynamics of the ridge. The shock velocity varies as $h^{1/2}$ (where $h \approx$ $\textbf{Ca}^{2/3}$ is the deposited film thickness as given by Bretherton), suggesting that the hydraulic jump physics is governed by a non-dimensional parameter similar to the Froude number $\textbf{Fr}$ but having the tube radius as a characteristic length, not the capillary length.\\ 

To conclude, present experiments support the following facts:
\begin{enumerate}
\item A non-stagnant film thicker than Bretherton's prediction is deposited on the non-wetting surface, due to the presence of a moving contact line.
\item A hydraulic jump (shock) is present in the film once a cutoff capillary number $\textbf{Ca}^*$ is exceeded.
\item The shock criterion is a non-dimensional parameter similar to $\textbf{Fr}$, which results when the capillary length is replaced with the tube radius (which is indeed the characteristic length).
\item The shock is caused by a capillary drainage mechanism made possible through the ridge dynamics.
\item Taylor bubbles form during film deposition, their length is in direct relation to the contact line and ridge dynamics.

\end{enumerate}


\bibliography{refs}

\bibliographystyle{unsrtnat}


\section*{Acknowledgements}
This work was partially supported by NSF CBET-0748049, DOE DE-FG36-07GO17018 and by the Department of Mechanical Engineering - Engineering Mechanics at Michigan Tech University.

\end{document}